# Mobile Food Printing in Professional Kitchens: An inquiry of potential use cases with novice chefs

Y. Kocaman, T. U. Bulut, O. Özcan

[1] *Koç University, KUAR Research Center for Creative Industries, Istanbul, Turkey.*

ABSTRACT: The knowledge transfer from 3D printing technology paved the way for unlocking the innovative potential of 3D Food Printing (3DFP) technology. However, this technology-oriented approach neglects user-derived issues that could be addressed with advancements in 3DFP technology. To explore potential new features and application areas for 3DFP technology, we created the Mobile Food Printer (MFP) prototype. We collected insights from novice chefs for MFP in the restaurant context through four online focus group sessions (N=12). Our results revealed how MFP can be applied in the current kitchen routines (preparation, serving, and eating) and introduce novel dining experiences. We discuss our learnings under two themes: 1) dealing with the kitchen rush and 2) streamlining workflows in the kitchen. The opportunities we present in this study act as a starting point for HCI and HFI researchers and encourage them to implement mobility in 3DFP with a user-oriented lens. We further provide a ground for future research to uncover potentials for advancing 3DFP technology.

## 1 INTRODUCTION

The competitive landscape forces restaurants to constantly seek innovative solutions to address food waste, menu limitations, and labor costs. 3D Food Printing (3DFP) technology is found to have the potential to help such challenges by automating tasks, especially laborious ones, in the restaurant kitchen (Pereira et al., 2022). Moreover, such digital fabrication tools elevate chefs' creativity by allowing a medium for new recipes (Mizrahi et al., 2016). These factors influence hospitality managers to implement robotic technologies to tackle competitive pressure and achieve relative advantage (Pizam et al., 2022). Despite its potential, 3DFP is still a technology that needs new features and capabilities to fully adapt (J. Y. Zhang et al., 2022).

The evolution of 3DFP technology can be traced back to 3D printing, which established its foundational design principles. The current trajectory of the 3DFP development primarily focuses on implementing features from 3D printing, evident in the shared design features between the two technologies (e.g., device proportions, nozzle diameters, chassis-based bodies, and axial constraints). Several studies work on advancing the food formulation and printing process to increase the usable ingredients and achieve the best textures and consistency at the end of the printing (Pereira et al., 2022). While this approach advances the mechanical capabilities of the technology, it reduces the food experience to the variety of ingredients and recipe precision. These studies formed an initial ground for the applicability of 3DFP to the food service sector; however, they neglect the user-derived issues in the kitchen that the technology could address (Deng et al., 2021, 2022). Although the knowledge transfer from 3D printing paved the way for unlocking the innovative potential of 3DFP technology, recent research asserts the importance of considering users' current practices, knowledge base, habits, and values when designing Human-Food Interaction (HFI) technologies given the food's unique nature as a material (Dolejšová et al., 2020). Overall, explorative research is needed to seek new features and application areas for 3DFP technology.

To respond to this inquiry, we aim to uncover use cases for 3DFP technology rooted in the situated practices in professional kitchens. We adopted the Research through Design (RtD) methodology and created the Mobile Food Printer (MFP) prototype. MFP prototype implements "mobility" as a novel feature for 3DFP. Using this prototype as a research object, we explored application areas in the restaurant context that mobility enables. We collected reflections on using MFP in the current routines of professional kitchens (preparation, serving, and eating) by conducting four online focus group sessions with 12 novice chefs. Moreover, we uncovered ideas for how such a novel technology might offer novel experiences for diners. We discuss what we learned from this process under two themes: 1) operability by diners or servers and 2) streamlining plating routines.

Our inquiry showed that MFP may help address conventional practices to streamline food preparation and serving in the kitchen. MFP also showed the potential to enhance diners' experience by facilitating playful eating. This transformation is uncovered by the user-centered approach we adopted while implementing mobility from 3D printing. By doing so, we were able to delve deep into the features of mobility and propose design insights for 3DFP devices specialized to the restaurant context.

## 2 RELATED WORKS

### 2.1 Mobility in 3D printing

Innovative applications in 3D printing have helped to unlock new possibilities for the technology. For instance, swarm printing concepts, often operated with materials such as plastic and concrete (Hunt et al., 2014; Oxman et al., 2014), opened up opportunities for building larger structures using this manufacturing method. In those examples, the outcome is built in collaboration with many synchronized small robots. With the advantage of free movement, the outcomes reach great sizes. In construction, it is emphasized that mobility increases the maneuver capability of the printer rather than stationary robotic examples. Increased maneuver capability is considered a novel advantage in small construction sites. Especially when mobility is combined with a robotic arm that enables the system to handle large load capacities while allowing it to work accurately (Keating et al., 2014). Even though some examples tested mobility with robotic construction (Hunt et al., 2014; Keating et al., 2014; X. Zhang et al., 2018), we haven't encountered a study that deeply explored mobility in digital food fabrication.

### 2.2 Adapting Mobility to Food Printing

Similar to most conventional types of 3D printers, most food printers are not mobile or portable devices that can be carried around. Only one product on the market offers portable use (*byFlow*, 2015). Portability allows this device to be used in novel dining concepts such as FOOD INK's pop-up dining concept (*3DFP Ventures Ltd.*, 2016). At FOOD INK, food is prepared by 3DFP devices at the sight of the diners. Although a portable food printer is commercially available (*byFlow*, 2015), mobility in food printing is still explored. Since portability extended the usage areas of the food printers, we speculate that if being mobile, contrary to being stationary, is possible in food fabrication, such devices could extend the potential use cases of 3DFPs.

As we also discussed above with examples, utilizing mobility in 3D printers is said to help overcome size limitations in architecture (Oxman et al., 2014). The mobility can also offer advantages to food printing, given that it increases the maximum printing size and maneuver capability inside the kitchen. In fact, when the dynamic nature of the kitchen and the dining room contexts are considered, mobility may entail more unique use cases in the kitchen and the dining room.

## 3 DESIGN AND IMPLEMENTATION OF MOBILE FOOD PRINTING (MFP) PROTOTYPE

We initially intended to design a robot to print edible materials by moving on varying surfaces. We further aimed to open up possibilities for mobility to pave the way for 3DFP technology. While building this prototype, we especially targeted free movement, which ultimately relates to responding to the current context in professional kitchens. In principle, we made a design proposal for a digital food fabrication device that prints edible materials on an indefinite surface. (Fig 1)

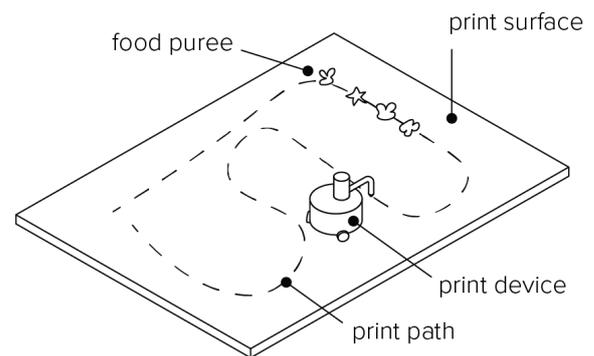

Figure 1: İnitial drawing defining the basic principles of the Mobile Food Printer (MFP) prototype

As this prototype aims to facilitate the exploration of new use cases and the design is prone to revisions, we built a low-fidelity prototype without depending heavily on resources. Instead, we collected available components online and designed parts necessary for the assembly and basic operations. We divided the MFP system into three main components: 1) Effector, 2) Arm, and 3) Extruder. (Fig. 2).

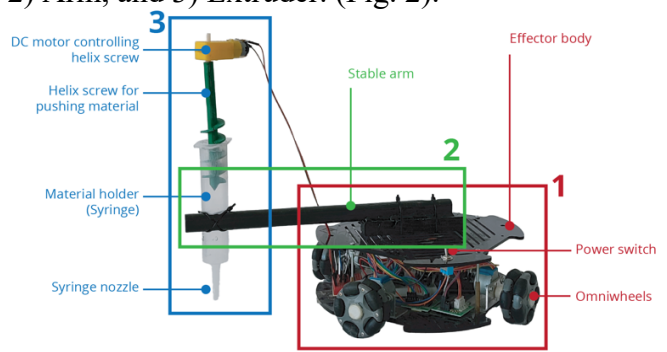

Figure 2: Components of Mobile Food Printer (MFP) conceptual prototype.

Effector: This component is primarily responsible for the movement of the printer. Thus, it required actuator motors and wheels that allow free movement on varying surface types and sizes. Additionally, this component highlighted the necessity for sensors to assist in navigating the printer on the printing surface. To achieve these functions, we used a premade robotic platform with four Omni wheels capable of moving freely on the x and y axes. The platform is controlled by an Arduino UNO processor, and each omni-wheel is driven by 28 BYJ-48 Geared Stepper Motor and ULN2003A Stepper Motor Driver circuit.

Arm: This component positions the extruder away from the platform to prevent overlaps during printing. In the initial design, we used a stable stick with future aims to give it the ability to move on the z-axis. We fixed the arm to the body with zip ties.

Extruder: In this part, we used 60 ml medical syringes as accessible food-safe parts that could be quickly supplied and replaced. We designed a 3D-printed helix screw to push the material from the syringe nozzle using Polylactic Acid (PLA). We attached a DC motor to the tip of the helix screw.

After we designed and assembled the components of the prototype, we made trials for printing basic geometrical shapes and complex models using our prototype. We first generated the G-code representation of a digital model by using the slicer software Inkscape and transferred the code to the Arduino UNO through a USB cable. Then, we prepared whipped cream and filled the syringe. After activating the code, the system succeeded in printing the geometries we intended to (Fig. 2).

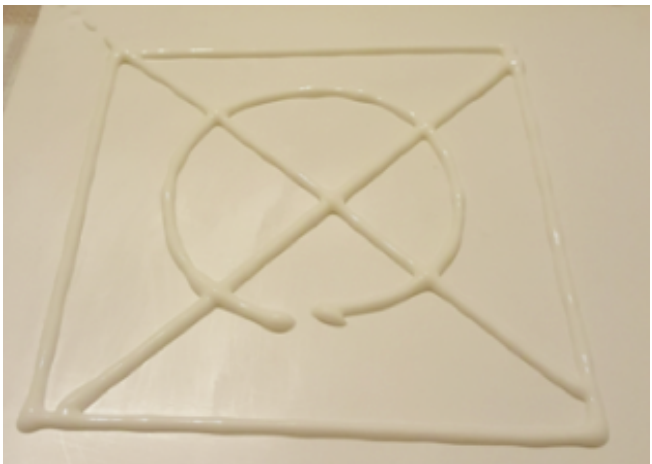

Figure 3: Initial test shape generated with MFP prototype.

The current degree of movement in the Z axis did not enable MFP to print complex 3D models. Thus, in line with our aims, we decided to pause the prototype development when we reached a lo-fi representation of our concept (Rudd et al., 1996). At this point, we continued to collect insights on this speculative artifact before spending any more effort advancing it.

At this developmental stage of the MFP prototype, we defined three cases that differentiate MFP from other 3DFP devices: 1) Printing on a large piece of food on a rectangular surface; 2) printing multiple small foods on a large rectangular surface; 3) printing food on different plates located on a table. We created video sketches illustrating how MFP operates to complete the tasks in these cases. (Fig 3)

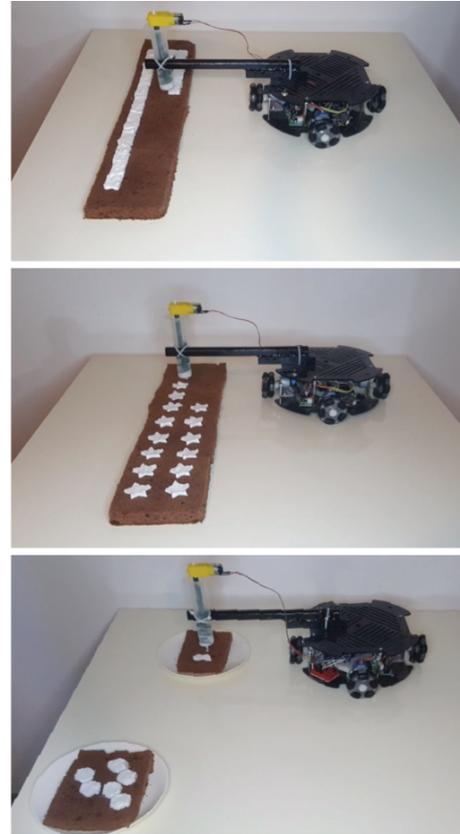

Figure 4: Screenshots from prototype demonstration videos. From top to bottom: 1) Printing on a large piece of food on a surface; 2) printing multiple small foods on a large surface; 3) printing food on different plates.

## 4 EVALUATION OF MFP: THE CASE OF NOVICE CHEFS

To explore potential discussions mediated by a 3DFP device design, we adopted the Research through Design (RtD) methodology and created the Mobile Food Printer (MFP) prototype (see section 3). With this prototype, we aim to trigger meaningful discussions about what the future of 3DFP could and should be (Olson & Kellogg, 2014, p. 169). Therefore, the knowledge we aim to produce within the scope of this study functions as proposals for the future, not predictions (Zimmerman et al., 2010).

A common practice in the Human-Food Interaction field is to collect experienced chefs' views while developing interactive kitchen technologies due to their advanced knowledge and experience working and managing professional kitchens (Genç et al., 2021; Mizrahi et al., 2016). However, a recent study highlights the significant difference in opinions on kitchen design among kitchen workers with changing experience levels (Teyin & Seçim, 2023). Thus, we focused on collecting opinions from a relatively underexplored community of novice chefs who are currently in or fresh out of culinary school. We anticipate that novice chefs potentially present an open mindset to the practices in professional kitchens. We reached twelve novice chefs through word-of-mouth in our professional circles and snowballing (Coskun et al., 2023). All our participants have experience working in professional kitchens as sous-chefs or interns between four months and sixty months ($M_{experience}$=21.75 months). (See Table 1 for details on participant characteristics). Two participants were professional food photographers trained in culinary arts (P8, P9). Three participants were pursuing double majors in nutrition and dietetics (P7, P10, P11).

Table 1. Novice chefs' participant characteristics.

| Session | Participant | Grade* | Work experience (months) |
|---|---|---|---|
| 1 | P1 | FG | 4 |
|   | P2 | FG | 24 |
|   | P3 | Senior | 12 |
| 2 | P4 | FG | 14 |
|   | P5 | FG | 35 |
|   | P6 | Senior | 60 |
| 3 | P7 | Senior | 6 |
|   | P8 | FG | 14 |
|   | P9 | FG | 14 |
| 4 | P10 | FG | 12 |
|   | P11 | FG | 60 |
|   | P12 | Senior | 6 |

* FG: fresh graduate.

In four online focus group sessions, we collected novice chefs' ideas on using MFP in their kitchen routines (preparation, serving, and eating) and how it might enhance their diners' overall experience. The sessions started with a brief presentation about the state-of-the-art of 3DFP technology and the current industry trends. We proceeded to introduce the MFP prototype and briefly explained how it works. After an initial exchange of perspectives on the prototype, we showed three cases that we speculated as potential competencies of MFP: 1) Printing on a large piece of food on a rectangular surface; 2) printing multiple small foods on a large rectangular surface; 3) printing food on different plates located on a table. (Fig. 3)

Each session lasted approximately one hour and was video recorded on Zoom. We then transcribed these recordings into text for analysis. We conducted thematic analysis (Braun & Clarke, 2006) on these transcriptions to reveal potential themes emerging from the dataset that could hint at future uses for MFP in professional kitchens.

## 5 RESULTS

We discovered potential advantages and relevant use cases for a Mobile Food Printer (MFP) to adapt to the commercial kitchen. Our results highlighted some aspects in particular to mobility, thus, could inspire specializing MFP to the restaurant kitchen.

### 5.1 Placing MFP in the dining room

In addition to the applications in the kitchen (i.e., backstage), our participants found the dining room (i.e., front stage) appropriate for elevating the presentation of tastes, visuals, and interactivity. Participants suggested extraordinary food presentations that could be enabled by the competencies of a mobile device.

*"I imagine roasting a whole turkey and putting some tiny garnish that would be almost impossible to do by hand. I think the contrast created could generate interest. (P1)"*

Here, P1 imagined a scenario where they prepare a huge spit roast in front of the diners and added tiny details using the MFP. They discussed how juxtaposing high-tech equipment (i.e., MFP) with ancient cooking techniques to capture diners' interest. They explained adding such a technological contrast to the dish design (a high-tech addition to a traditional recipe) would engage diners even more with the food they designed. They also discussed that adding sauces or decorations using state-of-the-art technology would render traditional recipes more captivating.

Learning 1: Using MFP as a presentation tool for food could foster novel dining experiences and ignite the attraction of diners to traditional dishes. Highlighting the technology and its new culinary capabilities could be used as a strategy to create such experiences.

Participants also envisioned practical concepts in the dining room where MFP enables overcoming size limitations. One participant suggested reducing the size of the MFP even more so that it could be attached to the corner of the dining table at the service of the diners. Just like in a Korean Barbecue, diners are envisioned as the primary agents operating the machine right at the dining table. By using this device, they would be able to add extra sauces and personalize their dish designs.

*"The device is located in the corner. Customers can add extra sauce using this bot when they run out of sauce on their plate. (P3)"*

They discussed that such a device would help track the diners' side dish consumption digitally. Such an application is mentioned to ease the tracking of the cost of foods by digitalizing the serving process. Our participants also envisioned a restaurant setting similar to an automated sushi bar, where food is delivered by a conveyor belt upon ordering. They envisioned that the device could help servers with the presentation of certain foods right at the table (e.g., serving foam right at the table). They further ideated future restaurant concepts that the MFP devices are the central agents that serve food. Participants noted that the restaurant's menu needs to be adapted for that particular serving style in such settings.

Learning 2: The dining experience could be enhanced right at the front stage as much as the backstage of the restaurant. Diners and food service staff are the potential agents of the device, and technology may act as the mediator of their interaction.

### 5.2 Dealing with the kitchen rush

The most frequent discussion point that arose during the sessions was how the kitchen staff could efficiently use MFP to reduce their overall workload. Our participants found MFP helpful in repetitive, time-consuming, and precision-intense tasks. They found the concept to aid the mise en place of the bakery station, which is often quite monotonous. However, the technology is found to be most beneficial for the plating. Here, participants elaborately described how the head chef needs to assign several kitchen staff to prepare plates for a dinner service of 100+ guests. In the current settings, participants noted that the kitchen staff assigned to such monotonous tasks get bored quickly and may do a sloppy job. They highlighted that a technology such as MFP could help reduce human error in such scenarios.

*"If we prepare a meal for five hundred people, all the plates come out simultaneously. Let's say five people are working there. One of them puts the meat, and the other squeezes the sauce. Someone else puts the decor next to it. There is a system that works like this. Using a mobile device that roams on the counter can be logical for that setting. (P9)"*

While a convenient 3DFP could aid the plating routine, mobility enriches this task with its ability to print on separate plates and move on the kitchen counter. It acts as a moving agent of cooking, just as any other kitchen staff operating monotonous tasks.

Learning 3: Mobility could help reduce manual labor and human error in the plating routines. Mobility also augments the role of the equipment as an agent of cooking.

### 5.3 Fitting a new machine into the kitchen space

Participants pressingly discussed that the kitchen space in restaurants must be used in the most efficient way possible, especially during the dinner service hustle. An additional device would require spare space, which is often not preferred as the current settings in *a la carte* kitchens already operate in small spaces. Participants offered several perspectives on how to reduce the space that MFP would occupy in the kitchen. First, they suggested that this device be adaptable to different food preparation stations (e.g., bakery and grill). This would help reduce the machine's storage need by scheduling the operation hours between stations.

*"The device should not be exclusive to stations. Since it is a portable device, every station can benefit from it. For example, it can also be used by the patisserie station for astonishing visuals and by a saucier to prepare thick sauces. (P1)"*

Second, they suggested transferring the device to move on the floor, thus not occupying counter space during plating. Lastly, they suggested designing the device to be able to be attached to a corner of the counter, therefore reducing the extra space it occupies on the counter. They further suggested adding a conveyor belt-like plating surface for extra space efficiency. Participants' first two suggestions assigned mobility to the device, while in the third suggestion, they fixed the device on the counter and assigned mobility to the remaining kitchen equipment (e.g., plates).

Learning 4: Mobility can be an advantage in improving space efficiency, especially counter space, when used at different food preparation stations. However, the objects to which mobility is assigned affect how the technology supports food preparation in the professional kitchen.

## 6 DISCUSSION

We adopted a user-centered approach while implementing mobility from 3D printing for developing food-experiencing artifacts. In doing so, we were able to delve deep into use cases for implementing the mobility feature to 3D Food Printing (3DFP) in the restaurant context. Overall, MFP demonstrated how to reshape a technology-driven research agenda when informed by the current routines and practices of the user (Deng et al., 2022). Our findings reveal initial insights into the design space of MFP based on culinary students' visions of collaborating with MFP in their daily work routines preparing and serving food.

### 6.1 Determining the end-user in the dining room

In contrast to the professional kitchen, which is a highly utilitarian space, the expected user experience of a restaurant's dining room is exceptionally hedonic. Although the prototype we developed in the scope of this research is not capable of operating advanced cooking tasks, it triggered ideas for potential applications in the dining room.

Numerous applications of interactive technologies augment the dining experience by developing novel and playful foods (Bonacho et al., 2020, p. 130; Deng et al., 2023) or novel eating scenarios with the help of technologies such as Augmented Reality (AR) (Batat, 2021). The gadget loving propensity of the consumers impacts the perceived value of the interactive restaurant self-service technologies. Putting high-tech innovation at the forefront of the food presentation and dining experience might capture the interest of such user types (Ahn & Seo, 2018). Dining experiences such as the FOOD INK, a pop-up interactive dining experience, mainly target these user groups. Such concepts offer entertaining and visually captivating dining experiences by making the 3DFP technology visible (*3DFP Ventures Ltd.*, 2016). This approach paves the way for the wider adoption of 3DFP technology and highlights the positive emotions enabled by putting the technology at the forefront of food preparation. Aligned with this course, our study participants envisioned taking advantage of the positive influence 3DFP creates on the diner. This was found especially critical for elevating the attention towards traditional foods. Chefs may develop improved food presentations using MFP and positively influence how diners attribute quality and perceive the value of traditional dishes.

A previous study investigating the integration of AR to positively affect customer experience suggests that restaurants should integrate novel technologies in stages depending on their customer profiles (Batat, 2021). In our case, mobility ignited proposals for adding new elements to the dining experience by uncovering the potential of treating the diner as the end-user of the technology. This finding suggests that an initial participant base needs to be identified who would be interested in using 3DFP as part of their dining out experience.

Previous research attempts to investigate the adoption of 3DFP technology and identifies potential user characteristics for interested domestic users (Brunner et al., 2018; Gayler et al., 2018; Kocaman et al., 2022). Adding to this body of knowledge, we show that using 3DFP as a diner in a restaurant is integrally different than using the technology in domestic settings. Although the user profiles may intersect, our findings highlight the importance of customer segmentation for the adoption of 3DFP in the restaurant context.

### 6.2 *Streamlining workflows in the kitchen*

The kitchen is a labor-dependent space, and the kitchen staff are the main agents who spend this labor. They are the grandmasters of the methods in the kitchen. The kitchen is a space with strict design guidelines, and when new interactive technologies enter this space, careful consideration needs to be made of how this new technology will affect the current kitchen layout (Claus Bech-Danielsen, 2012). Suggesting where the new technology will fit in the kitchen layout is a crucial point that requires recognition. Interactive kitchen technologies developed for extreme settings such as space missions have strict design requirements. For instance, space foods must be lightweight and easy to store (Enfield et al., 2023). To maximize the use of space, the design for such extreme environments prioritizes practical features.

In our case, we observed several design possibilities for MFP that do not require a fixed place for the device (a detachable add-on to the service counter, used between countertops). MFP showed potential applications for reducing the human labor in the plating due to its high maneuver capability and taking little counter space. Moreover, the technology could be treated as a kitchen staff member who could undertake monotonous tasks and reduce errors. These applications allowed by mobility suggest increasing the agency of technology in food-making. Thus, suggests leveraging the agency of technology over the persons to streamline the workflows in professional kitchens (Altarriba Bertran et al., 2019).

## 7 CONCLUSION AND FUTURE WORKS

In conclusion, this study delves into the design of Mobile Food Printer (MFP) technology, seeking to explore new features and application areas for 3D Food Printing (3DFP) technology. We specifically investigated the implementation of the mobility feature and discovered that MFP might digitize food serving and foster increased interactivity between diners, chefs, and the kitchen staff. We acknowledge the limited sample size of this study, however, due to our exploratory approach, we were able to uncover fruitful potentials for the adoption of 3DFP technology in the restaurant context. To our knowledge, this is the first study that underlines these potentials, particularly enabled by the feature of mobility. Uncovering these potentials helps foresee alternative futures with 3DFP technology. We observed that 3DFP technology has the potential to enhance users' eating experiences in the restaurant context. The set of proposals we present in this study acts as a starting point for HCI and HFI researchers and encourages them to implement new features to 3DFP with a user-oriented lens.

Our results form a basis for future research to dig deeper into the feature of mobility and its potential to create curiosity for other potential features that might be explored in the food printing domain (e.g., drone printing). In the following steps, we aim first to reflect the user insights onto the prototype and test it in the actual kitchen settings.